\documentclass[12pt]{article}

\usepackage{graphicx,float,subfigure}
\usepackage{epsf,amsmath,bbold,amsfonts}
\usepackage{appendix}
\usepackage{amssymb}

\usepackage{paper2e}

\usepackage{xcolor}
\usepackage{hyperref}
\hypersetup{colorlinks=false,pdfborderstyle={/S/U/W 1}}

\def\a{\alpha}

\def\c{\chi}

\def\d{\delta}
\def\D{\Delta}

\def\eps{\varepsilon}
\def\f{\frac}

\def\G{\Gamma}

\def\mc{\mathcal}

\def\m{\mu}
\def\n{\nu}

\def\nn{\nonumber}

\def\p{\partial}

\def\s{\sigma}

\def\be{\begin{equation}}
\def\ee{\end{equation}}

\def\bea{\begin{eqnarray}}
\def\eea{\end{eqnarray}}

\def\ba{\begin{array}}
\def\ea{\end{array}}

\def\bc{\begin{center}}
\def\ec{\end{center}}

\def\bl{\begin{flushleft}}
\def\el{\end{flushleft}}

\def\br{\begin{flushright}}
\def\er{\end{flushright}}

\def\bi{\begin{itemize}}
\def\ei{\end{itemize}}

\def\bt{\begin{tabular}}
\def\et{\end{tabular}}

\begin{document}

\begin{titlepage}
\vspace{5cm}

\vspace{2cm}

\begin{center}
\bf \Large{Dilaton: Saving Conformal Symmetry}

\end{center}

\begin{center}
{\textsc {Frederic Gretsch, Alexander Monin}}
\end{center}

\begin{center}
{\it Institut de Th\'eorie des Ph\'enom\`enes Physiques, \\
\'Ecole Polytechnique F\'ed\'erale de Lausanne, \\ 
CH-1015, Lausanne, Switzerland}
\end{center}

\begin{center}
\texttt{\small frederic.gretsch@epfl.ch} \\
\texttt{\small alexander.monin@epfl.ch} 
\end{center}

\vspace{2cm}

\begin{abstract}

The characteristic feature of the spontaneous symmetry breaking is the presence of the Goldstone mode(s). For the conformal symmetry broken spontaneously the corresponding Goldstone boson is the dilaton. Coupling an arbitrary system to the dilaton in a consistent (with quantum corrections) way has certain difficulties due to the trace anomaly. In this paper we present the approach allowing for an arbitrary system without the gravitational anomaly to keep the dilaton massless at all orders in perturbation theory, i.e. to build a theory with conformal symmetry broken spontaneously.

\end{abstract}

\end{titlepage}

\newpage
\section{Introduction and motivation}

One of the big conceptual questions that has yet to be answered is the status of naturalness in the case of the Higgs mass, or in other words the fine-tuning problem\footnote{For a recent perspective see for example~\cite{Shaposhnikov:2007nj,Giudice:2008bi,Giudice:2013yca}.}. The solution to the problem having its roots in the naturalness principle~\cite{'tHooft:1979bh} and the way we describe the world might lead to a new paradigm.

The idea of neglecting power divergences\footnote{Quadratic divergence is not the culprit {\it per se}, since the cutoff can be considered as being unphysical. The problem should be formulated in terms of physical scales, say mass of a new heavy particle. We use the expression quadratic (or in general powerlike) divergence as a jargon simply for the brevity.} put forward some time ago~\cite{Bardeen:1995kv,Meissner:2006zh,Meissner:2007xv,Shaposhnikov:2008xi} finds its way in the light of the new LHC data to the reformulation of the naturalness principle~\cite{Farina:2013mla,Ghilencea:2013fka,Craig:2013xia,Dubovsky:2013ira,Aoki:2012xs}. The reason for neglecting all power divergences can be the scale invariance of the system and it can help in understanding the hierarchy problem~\cite{Shaposhnikov:2007nj,Shaposhnikov:2008xi,Shaposhnikov:2008xb}. Sometimes the presence of the scale invariance at the classical level is considered to be sufficient for solving the problem of naturalness~\cite{Meissner:2006zh,Meissner:2007xv,Nishino:2007mj,Foot:2007iy,Hambye:2013dgv,Oda:2013cpa,Heikinheimo:2013fta,Orikasa:2013ypa,Kobakhidze:2013uoa,Oda:2013rx,Nishino:2004kb}. We believe that -- if at all -- it should be the symmetry of the quantum system but not of the classical one only. Therefore, it is crucial to know whether a system can be extended to have the scale symmetry.

The Standard Model (SM) is classically scale invariant if the Higgs mass is put to zero. The way to make the Lagrangian of the SM scale invariant even for a non-zero Higgs mass is well known~\cite{Coleman_aspects}. In this case the mass originates from a vev of a new dynamical field -- the dilaton. In general a classical symmetry group of a system can be extended by adding corresponding Goldstone particles, {\it i.e.} considering nonlinear realization of the group~\cite{Coleman:1969sm,Callan:1969sn,Salam:1969rq,Salam:1970qk}. However, this procedure does not guarantee that once the quantum corrections are taken into account the symmetry stays intact. The reason is that sometimes quantum anomalies appear revealing non-invariance of the system. Technically the origin of anomalies is the regularization/renormalization procedure. It is said that the symmetry is not consistent with a regulator. In the case of the scale symmetry, for example, cut-off, Pauli-Villars and dimensional regularizations introducing a mass scale break the symmetry explicitly.

Therefore, if a regularization preserving the scale invariance is found one can expect that the quantum system stays scale invariant. Such a regularization was proposed in~\cite{Shaposhnikov:2008xi}. It is reminiscent of the one discussed in~\cite{Englert:1976ep}. The idea of the regularization is that in order to make the symmetry manifest all the scales (including the one coming from a regulator) have to take their origin from the vev of the dilaton. The scale invariance of the system then follows from the dimensional analysis~\cite{Shaposhnikov:2009nk}.


Scale and conformal symmetries can be used interchangeably provided a certain condition is met~\cite{Coleman_aspects,Callan:1970ze,DiFrancesco:1997nk}. However, whether the prior implies the latter in general is still debated~\cite{Fortin:2012hn,Luty:2012ww}. Therefore, a possibility of preserving the conformal invariance at the quantum level needs a separate investigation. The scheme proposed in~\cite{Shaposhnikov:2008xi} allows to introduce a regulator in a conformally invariant way. In~\cite{Armillis:2013wya} it was shown that for such a regularization in a certain class of models the conformal invariance
is indeed preserved at the one loop level. In other words it was shown that no anomaly appears in the trace of the energy momentum (EM) tensor and it stays zero at one loop.

Having a regularization respecting the symmetry is not the end of the story though. Simple dimensional analysis does not allow to prove that renormalized theory is conformally invariant, i.e. that the counter terms needed to cancel divergences are conformal at any order in perturbation theory. The reason is that the two operations the renormalization and the symmetry transformation do not necessarily commute. The well known example is the Weyl anomaly which in dimensional regularization comes from the non-invariance of the counterterms.

Let us be more specific and demonstrate what we mean. We consider a theory given by a classical action $S[\phi]$. A given transformation $T _s$
\be
\phi \to \phi + T_s ( \phi )
\label{sym_trans_gen}
\ee
is called a symmetry of the classical action if
\be
T _ s ( S[\phi] ) = 0.
\ee
At the quantum level the symmetry is manifested through the Ward identities. One has to check that adding the counterterms does not spoil the symmetry. In other words all the counter terms have to be invariant under the transformation (\ref{sym_trans_gen}). The counter terms can be found from the regularized effective action $\G [\phi]$ which is the Legendre transform of $W[J_\phi]$, defined as
\be
e ^ {i W[J _ \phi ]} = \int \mathcal{D} \d \phi \, e ^ {i S[\phi + \d \phi] + i \int J _ \phi \d \phi }.
\ee

In dimensional regularization one considers the system in $n=4 - 2 \eps$ dimensional space. The effective action in general has the form
\be
\G [\phi] = \G _ P [\phi] + \G _ F [\phi],
\ee 
where $\G _ P [\phi]$ contains poles in $\eps$, while $\G _ F [\phi]$ is a finite part. Now if the measure and the regularized action are invariant under (\ref{sym_trans_gen}) it is clear that
\be
T_ s (\G [\phi]) = 0.
\ee
However that does not mean that both the pole and finite parts of the effective action are automatically invariant under the symmetry (\ref{sym_trans_gen})
\be
T _ s ( \G _ P [\phi] ) = T _ s ( \G _ F [\phi] ) \overset {?} {=} 0.
\ee
It is only true if the symmetry transformation $T_s$ does not change the $\eps$ behavior of $\G_{P,F}$. In other words the variation of the pole part $\G _ P [\phi]$ does not acquire finite in $\eps$ piece, so that it does not "mix" with $\G _ F [\phi]$. For symmetries which do not depend on the number of dimensions explicitly the condition is obviously satisfied
\be
T _s (\G _ P [\phi]) = \G' _ P[ \phi],
\ee
and the counter terms are also invariant. Hence, in this case it is enough to have a symmetry preserving regularization in order to conclude that the symmetry is preserved at the quantum level.

It is not so for symmetries depending on the number of space-time dimensions. For example in the case of the Weyl symmetry although the transformation of the metric tensor itself is $n$-independent
\be
g ' _ {\m \nu} (x) = e ^ {2 \s} g  _ {\m \nu} (x),
\ee
its determinant $\det g _ {\m \n} \equiv g $ acquires $n$-dependent piece
\be
g' (x) = e ^ {2 n \s} g (x).
\ee
The effective action for the metric considered as a source contains a term
\be
W[g _ {\m \nu}] \supset  \f {a} {\eps} \int d ^ n x \sqrt {-g} \, {E _4},
\ee
with $E _ 4$ being the Euler density in $4$ dimensions, which gives rise to the $a$ anomaly~\cite{Capper:1974ic,Duff:1993wm,Deser:1993yx}.

So far we have seen that for symmetries that explicitly depend on the number of dimensions -- which is the case for the special conformal transformations (see below) -- invariance of the renormalized theory does not follow immediately from the fact that the theory can be regularized in the symmetry preserving way. It does not mean though that the symmetry is lost. One may rearrange the terms in $\G[\phi]$ in such a way that they transform under the conformal transformations independently. 

This is precisely what we do in the present paper. We focus our attention on systems without the gravitational (Diff) anomalies~\cite{AlvarezGaume:1983ig,AlvarezGaume:1985vx}. We show that for a Diff anomaly free theory the classical conformal invariance can be promoted to the quantum symmetry at all orders in perturbation theory. We also perform the renormalization of two toy models as a consistence check. We show by explicit computations that the three loop counter terms can be chosen in a conformally invariant form.

The paper is organized as follows. In the Section \ref{gen_arg} we present a general proof of a possibility to preserve the conformal symmetry at the quantum level. Explicit computations for two toy models are made in the Section \ref{exp_com} to demonstrate how the procedure works. We give our conclusions in Section \ref{conc}. Relevant 3-loop graphs with corresponding leading divergences are given in the Appendix \ref{app_A}.

\section{General argument \label{gen_arg}}

Here we are going to prove that the conformal invariance can be promoted to the quantum symmetry by means of adding the dilaton. The idea of promoting all mass scales to the dynamical field can be used in general for any known regularization. However, for our purposes to simplify computations we use dimensional regularization. For an alternative regularization see~\cite{Shaposhnikov:2008ar}. Another simplification comes from dealing only with scalar fields. The extension to spinor and vector fields is straightforward.
We present an iterative procedure allowing to perform the renormalization in a conformally invariant way at all orders in perturbation theory.

The general setup is as follows. We consider a theory described by a conformally invariant $4$-dimensional classical action~$S [\Phi]$. Within the dimensional regularization some couplings in the regularized action become dimensionful. We introduce the dilaton field $X$ with the canonical kinetic term to compensate for the non-invariance of the action with respect to the conformal transformations in $n$ dimensions. As a result we have conformally invariant regularized action $S[\Phi,X]$. Now we start the iterative procedure. The regularized $1$-loop effective action for the new system can be written in the following form
\be
\G [\Phi,X] = \G _ P [\Phi,X] + \G _ F [\Phi,X].
\ee
The scale invariance of both $\G _ {P,F}$ is manifest. Simple dimensional analysis proves it. Therefore, the scale invariance is preserved. At the same time since the special conformal transformations depend on the number of dimensions conformal invariance is not evident already at one loop.

To make a definitive conclusion about the conformal invariance we somewhat modify our approach, using the one from~\cite{Luty:2012ww}. Instead of computing the effective action in flat space time we compute it in a curved background $g _ {\m \n}$, taking the metric to be non-dynamical. We assume that since the regularized action is conformally invariant it is possible (see~\cite{Iorio:1996ad}) to couple the system to gravity $ S[\Phi, X, g _ {\m \n}] $ respecting the Weyl transformations
\bea
\Phi (x) & \to & e ^ {- \s \D } \Phi (x), \nn \\
X (x) & \to & e ^ {- \s \D } X (x), \nn \\
g _ {\m \nu} (x) & \to & e ^ {2 \s} g  _ {\m \nu} (x),
\label{Weyl_transform}
\eea
where $\D = {n} / {2} - 1$ is the scaling dimension for both $\Phi$ and $X$.

Then the $1$-loop effective action can be computed from
\bea
\G [\Phi, X, g _ {\m \n}] & = & W [J _ \Phi, J _ X, g _ {\m \n}] - \int d ^ n x \sqrt {-g} \Phi J _ \Phi - \int d ^ n x \sqrt {-g} X J _ X,
\nn \\
e ^ {i W[J _ \Phi, J _ X, g _ {\m \n}]} & = & \int (\mathcal{D} \d \Phi) \, (\mathcal{D} \d X)
e ^ {i S[\Phi + \d \Phi, X + \d X, g _ {\m \n}] + i \int \sqrt{-g} J _ \Phi \d \Phi + i \int \sqrt{-g} J _ X \d X},
\eea
where the scaling dimensions of the sources $J_{\Phi,X}$ are chosen in such a way as to preserve the Weyl symmetry. 
It should be stressed that the regularized (not renormalized yet) $1$-loop effective action
\be
\G [\Phi, X, g _ {\m \n}] = \G _P [\Phi, X, g _ {\m \n}] + \G _ F [\Phi, X, g _ {\m \n}]
\ee
is invariant under the Weyl transformation, while divergent and finite parts separately are not in general.

We expect the conformal symmetry to be broken spontaneously, therefore, the dilaton has a non-zero vev. Hence, X can be chosen in the form
\be
X = v e ^ { \tau \D}.
\ee
Performing the Weyl transformations (\ref{Weyl_transform}) with $\s = \tau$
\bea
\hat \Phi (x) & = & e ^ {- \tau \D} \Phi (x), \nn \\
\hat g  _ {\m \nu} (x) & = & e ^ {2 \tau (x)} g  _ {\m \nu} (x), \nn \\
\hat X & = & v,
\eea
we get as a result
\be
\G [\Phi, X, g _ {\m \n}] = \G [\hat \Phi, v, \hat g _ {\m \n}] = \G _ P [\hat \Phi, v, \hat g _ {\m \n}] 
+ \G _ F [\hat \Phi, v, \hat g _ {\m \n}].
\label{conf_eff}
\ee

Systems without gravitational anomalies are invariant under the general covariant transformations (under the diffeomorphisms).
The absence of the Diff anomaly is guaranteed in theories without chiral fermions. With those included certain relations between their charges have to be satisfied to cancel the anomaly. For example, the SM is Diff anomaly free~\cite{Peskin:1995ev}. Assuming that the theory at hand {\it does not} have the Diff anomaly we conclude that both terms in the effective action (\ref{conf_eff}) are Diff invariant. Hence, taking the background metric to be 
\be
\hat g _ {\lambda \s} (x)  = \eta _ {\lambda \s} e ^{2 \tau},
\ee
which corresponds to the flat initial metric
\be
g _ {\m \n} = \eta _ {\m \n},
\ee
we consider the general covariant transformation corresponding to the special conformal one
\be
x' _ \m =\f {x _ \m - a _ \m x ^ 2}{1 - 2 (a x) + a^2 x^2}.
\ee
In this case the fields transform in the standard way
\bea
\hat \Phi' (x') & = & \hat \Phi (x), \nn \\
\hat g' _ {\m \n} (x') & = &\hat g _ {\lambda \s} (x)\f {\p x ^ \lambda} {\p x ^ {' \m}} 
\f {\p x ^ \s} {\p x ^ {'\n}} = \eta _ {\m \n} e ^ {2 \tau (x)} \Omega ^ 2 (x),
\eea
where $\Omega (x) = 1 - 2 (a x) + a^2 x^2 $. On the other hand those transformations can be realized keeping the metric 
$g _ {\m \n} = \eta _ {\m \n} $ intact in the following form
\bea
e ^ {2 \tilde \tau (x')} & = & e ^ {2 \tau (x)} \Omega ^ 2 (x), \nn \\
\tilde\Phi (x') & = & \Omega ^ \D \Phi (x),
\eea
which is precisely the conformal transformation of the fields in flat space time~\cite{Coleman_aspects,DiFrancesco:1997nk,Blumenhagen:2009zz}. As a result the pole part of of the (\ref{conf_eff}) is conformally invariant. Thus the $1$-loop counter term can be chosen to be conformal. Adding the new counter term to the classical (tree) action and repeating the procedure one builds $2$-loop conformally invariant counter terms. Iteratively extending the procedure finalizes the proof.

\section{Examples \label{exp_com}}

In the previous section we presented the proof that the conformal symmetry can be preserved at the quantum level in theories without the Diff anomaly by introducing the dilaton provided the symmetry consistent regularization is employed. It is necessary to have several explicit computations corroborating the proof. In this section we consider two examples performing the renormalization and computing the counter terms to show that indeed the conformal invariance can be brought to the quantum level.

\subsection{Example in four dimensions}

As the first example let us consider the standard $\phi^4$ theory given by the Lagrangian\footnote{The index of the Lagrangian specifies the loop order.}
\be
\mc {L}_0 = \frac 1 2 \p_\nu\phi \p^\nu \phi - \dfrac {\lambda}{4!} \phi ^ 4.
\ee 
The theory is conformally invariant at the classical level in four dimensions. However, as soon as the dimensional regularization is used and the Lagrangian is extended to $n = 4 - 2 \eps$ dimensions the dilaton is needed to preserve the conformal symmetry
\be
\mc L_0 = \frac 1 2 \p_\nu\phi \p^\nu \phi + \dfrac 1 2 \p_\nu X \p^\nu X - \dfrac {\lambda}{4!}\phi^4 X^{2\alpha},
\label{examp_4d}
\ee
where $\a$ is chosen in such a way as to compensate for the non-invariance of the $\phi ^ 4$ term
\be
\alpha = \frac{4-n}{n-2}.
\label{def_alpha}
\ee
The way to deal with fractional power of $X$ was previously discussed in~\cite{Shaposhnikov:2008xi,Shaposhnikov:2009nk,Armillis:2013wya}. It basically amounts to consider a perturbative expansion around the vev $v$ of the dilaton
\be
X = v + \c,
\ee
which leads to the following expansion of the potential
\be
 \phi^4 X^{2\alpha} = \phi^4 v^{2\alpha} 
 \left [
 1 + 2 \eps \ln \left(1+\frac \chi v\right)+ 2\eps^2\left(\ln\left(1+\frac \chi v\right) + \ln^2\left(1+\frac \chi v\right) \right)+  O\left(\eps^3\right)
\right].
\label{potential_exp}
\ee

One technical remark should be made here. The dimensionality of the vev $v$ is not the one of the mass scale (for $n \neq 4$). Therefore, the analog of the renormalization scale $\m$ in dimensional regularization is played by a certain power of $v$. Namely,
defining $\mu$ as
\be
\m ^ {2 \eps} = v ^ {2 \a},
\ee
one has to keep it fixed when taking the limit $\eps \to 0$.

\begin{figure}[H]
\centering

\subfigure[$O(\eps^{-1})$]{
\label{fig:1_loop}
\includegraphics[width=3cm]{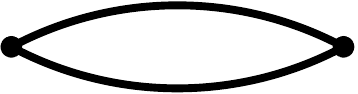}
}

\subfigure[$O(\eps^{-2})$]{
\label{fig:2_loop_II}
\includegraphics[width=3cm]{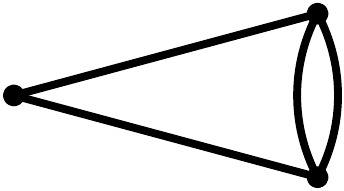}
}
\subfigure[$O(\eps^{-2})$]{
\label{fig:2_loop_I}
\includegraphics[width=3.5cm]{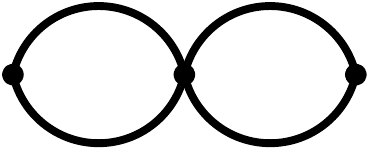}
}
\subfigure[$O(\eps^{-1})$]{
\label{fig:2_loop_prop}
\includegraphics[width=3cm]{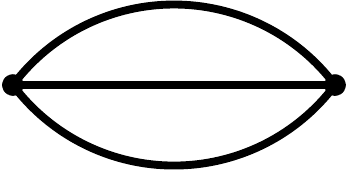}
}
\caption{\label{fig:1_2_loops_top}
One and two loop UV divergent topologies and their leading divergence.}
\end{figure}

At the one loop level only the first term contributes to the divergent graphs, because the others have at least one factor of 
$\eps$ which if taken into account renders any one-loop graph finite. Therefore, it is clear that at one loop the counter term is exactly the same as in $\phi ^4$ theory without the dilaton, which can be made conformal by adding finite renormalization. Namely, the renormalized theory is described by the one loop bare Lagrangian
\be
\mc L_1 = \mc L_0 - \frac{\lambda^2}{(4\pi)^2} \frac{3}{2\eps} \f {\phi ^ 4}{4!} X^{2\alpha}.
\ee
For two loops the counter terms have the form
\begin{equation}
  \mc L_2 = \mc L_1
  + \frac 1 2 \delta_{Z,2} \p_\nu\phi \p^\nu \phi - \dfrac {\delta_{\lambda,2} }{4!}\phi^4 X^{2\alpha}.
  \label{Lagrangian_2l} 
\end{equation}
Now the second term in (\ref{potential_exp}) has to be taken into account as well, bringing other interaction vertices.  
All of them have four $\phi$ and arbitrary many $\c$ legs. The topologies for the one and two loop divergent diagrams are presented in Fig.\ref{fig:1_2_loops_top}. None of the diagrams with $\c$ propagating in the loop is divergent for each internal $\c$ line brings at least a factor of $\eps^2$.

Thus, considering only $\phi$ external legs is equivalent to studying the $\phi^4$ theory. It is known~\cite{Kazakov:1979ik,Kleinert:1991rg,Peskin:1995ev} that the following choice of coefficients  in the Lagrangian (\ref{Lagrangian_2l})
 \begin{equation}
  \delta_{Z,2}=-\frac{\lambda^2}{(4\pi)^4}\frac{1}{24\eps}, \qquad
 {\label{eq-2L-ct}}
\delta_{\lambda,2}=\frac{\lambda^3}{(4\pi)^4}\frac 3 4 \left[\frac{3-2\eps}{\eps^2}\right].
\end{equation}
makes the $\phi^4$ theory finite. Let us introduce $\chi$ legs. According to Fig.\ref{fig:1_2_loops_top} and the $\eps$ expansion of the interaction Lagrangian (\ref{potential_exp}), divergences with $\chi$ legs can only subsist in diagrams of topologies Fig.\ref{fig:2_loop_I} and Fig.\ref{fig:2_loop_II}, but only if all these legs are connected to the same vertex. As a consequence at the two loop level there is no need for $X$ kinetic counter-term\footnote{Kinetic counter-term is associated to Fig.\ref{fig:2_loop_prop} topology.}. It also implies that the problem of divergences with $N$ $\chi$ legs is reduced to the combinatorics of all possible insertions of vertices with $\c$ field. It turns out that (\ref{eq-2L-ct}) is sufficient to cancel all divergences in two-loop diagrams.

Renormalization of the theory at hand at three loops is the first nontrivial check of the general proof, since the result for up to two loops can be expected based on~\cite{Luty:2012ww,Komargodski:2011vj,Komargodski:2011xv}, where the conformal invariance was proven for the non-dynamical dilaton. At three loops there are divergent diagrams containing the dilaton in the loop (see Appendix \ref{app_A}). A lengthy computation leads to
\be
 \mc L_3 =  \mc L_2 + \frac 1 2 \delta_{Z,3}\p_\nu\phi \p^\nu \phi 
  - \dfrac {\delta_{\lambda,3}}{4!}\phi^4 X^{2\alpha} 
  -\frac{C_6}{6!}\dfrac{\phi^6}{X^2}X^{2\alpha}
-\frac{C_8}{8!}\dfrac{\phi^8}{X^4}X^{2\alpha},
\label{Lagrangian_3l}
\ee
with
\begin{equation}
\delta_{Z,3}=-\frac{1}{4}\frac{\lambda^3}{(4\pi)^6}
\frac{1-\eps/2}{6\eps^2},
\qquad
 C_6 = \frac{\lambda^4}{(4\pi)^6} 180 \frac 1 \eps,
\qquad
C_8 = \frac{\lambda^4}{(4\pi)^6}  \frac{9625}{6}\frac 1 \eps.
\end{equation}
All the terms in the Lagrangian (\ref{Lagrangian_3l}) are obviously conformally invariant. As one can see the new counter terms are needed to account for all divergences at three loops, thus, making the theory non-renormalizable.

\subsection{Example in eight dimensions}

Although the results of the previous section are not trivial the counterterms are at most quadratic in derivatives. The terms with more derivatives are going to appear at higher loop order. Instead of continuing with the first toy model taking into account more and more loops, in this section we consider another toy model in $8$ space-time dimensions to illustrate that the counter terms with more than two derivatives can also be chosen to be conformal. In this model such counter terms appear already at one loop. The second toy model is given by the following Lagrangian
\begin{equation}
\mc L_0 = \frac 1 2 \p_\nu\phi \p^\nu \phi + \dfrac 1 2 \p_\nu X \p^\nu X - \dfrac {g}{3!}\phi^3 X^{1+2\alpha},
\label{lagr-n8}
\end{equation}
with $\a$ is defined in (\ref{def_alpha}) with the exception that now the number of dimensions is
\be
n = 8 - 2 \eps.
\ee
Expanding around the vev of the dilaton leads to
\begin{equation}\label{eq-Lagr-div-1L}
 \mc L_0 = \frac 1 2 \p_\nu\phi \p^\nu \phi + \dfrac 1 2 \p_\nu \chi \p^\nu \chi - \dfrac {g}{3!}\phi^3 v^{1+2\alpha}
\left[\left(
1 + \frac{\chi}{v}
\right)^{-\frac 1 3} + O(\eps)\right].
\end{equation}

At one loop there are only three different topologies for UV divergent graphs, the ones with two three and four vertices. These diagrams give rise to the counter terms with four, two and zero derivatives correspondingly
\begin{equation}
\mc L_1 = \mc L_0 - \delta^4- \delta^2 - \delta^0
\end{equation}
where 
\be
\delta^P=\sum\limits_N\delta^P_N,
\ee
and $\delta^P_N$ denotes conformally invariant counter term with $P$ derivatives and $N$ $\phi$ legs.
Straightforward computation shows that momentum dependent counter-terms can be chosen as 
\begin{equation}
  \begin{aligned}
  \delta^4_2
&=
\frac{g^2}{2\eps} \frac{1}{(4\pi)^4}\frac{1}{120}
&&{
\left[
\p^2\left(\phi X^{\frac{2}{2-n}}\right)
\right]^2,
}
\\
\delta^4_4 
&=
\frac{g^2}{2\eps} \frac{1}{(4\pi)^4}\frac{c_1^2}{60} 
&&{
\left[
\p^2\left(\phi^2 X^{\frac{n}{2-n}}\right)
\right]^2,
}
\\
 \delta^4_6
&=
\frac{g^2}{2\eps} \frac{1}{(4\pi)^4}\frac{c_2^2}{120}
&&{
\left[
\p^2\left(\phi^3 X^{\frac{2n-2}{2-n}}\right)
\right]^2,
}
  \end{aligned}
\end{equation}
and
\begin{equation}
\begin{aligned}
 \delta^2_3
 &=
\f {g^3} {\eps}  \frac{1}{12(4\pi)^4} \frac {1}{ 3!}
&&
\left\{
\phi^2 X^{-2/3}
\p^2(\phi X^{-1/3}) - \phi X^{-1/3}\left[\p_\mu( \phi X^{-1/3})
\right]^2
\right\},
\\
 \delta^2_5
 & = 
\f {g^3} {\eps} \frac{1}{12(4\pi)^4} \frac {c_1^2 }{2}
&&
\left\{
\phi^4 X^{-8/3}
\p^2(\phi X^{-1/3}) - \phi X^{-1/3}\left[\p_\mu( \phi^2 X^{-4/3})
\right]^2
\right\},
\\
 \delta^2_7
 &= 
\f {g^3} {\eps} \frac{1}{12(4\pi)^4}\frac {c_1^2 c_2 }{ 2}
&&
\left\{
\phi^4 X^{-8/3}
\p^2(\phi^3 X^{-7/3}) - \phi^3 X^{-7/3}\left[\p_\mu( \phi^2 X^{-4/3})
\right]^2
\right\},
\\
 \delta^2_9
 &
 =
\f {g^3} {\eps} \frac{1}{12(4\pi)^4} \frac {c_2^3}{ 3!}
&&
\left\{
\phi^6 X^{-14/3}
\p^2(\phi^3 X^{-7/3}) - \phi^3 X^{-7/3}\left[\p_\mu( \phi^3 X^{-7/3})
\right]^2
\right\},
\end{aligned}
\label{counter_term_8}
\end{equation}
where $c_k$ are the coefficients of the expansion
\be
\left(
1 + \frac{\chi}{v}
\right)^{-\frac 1 3} = 1 + \sum\limits_{k=1}^{+\infty} (-1)^k\frac{c_k}{k!} \left(\frac{\chi}{v}\right)^k.
\ee
Momentum independent counter-terms are trivially found to be 
\begin{equation}
 \delta^0_M = \frac {g^4 }{\eps} C_M \phi^M X^{2\alpha +4-M}\quad M=4,6,8,10,12
\end{equation}
where $C_M$ is an $\eps$ independent constant. One can easily check that all the counter terms are conformally invariant. Their structure can not be fixed by dimensional analysis solely, i.e. the scale invariance is not restrictive enough. For instance, both terms in curly braces in (\ref{counter_term_8}) are scale invariant separately, and they can be taken with arbitrary coefficients without spoiling the scale symmetry. Conformal invariance in turn is only present if the relative minus sigh is taken.

\section{Conclusion and discussion \label{conc}}

The question whether it is possible to extend the symmetry group of a quantum system to include the conformal transformations is very important. It can be interesting not only from the model building point of view or as a first step in understanding the mind boggling problem of fine tuning but also from a perspective of purely formal quantum field theory. 

In many cases the conformal invariance although present in the classical system is lost upon taking into account the quantum effects. The source of the trace anomaly is a hidden mass scale needed for a regularization. The symmetry gets broken already at the stage of regularizing the theory. Therefore, it is reasonable to look for a regularization that does not break the conformal invariance.

It is known~\cite{Coleman_aspects} how to make the scale invariance explicit at the classical level in a theory with mass parameters. Allowing those parameters to transform, compensating for the non-invariance under the scale transformations, makes the classical action scale symmetric. Similar idea is put forward in building quantum scale invariant theory. Taking the origin of all mass scales -- including the regularization related one -- being the vev of the dilaton leads to an obviously scale invariant theory.

In the case of conformal invariance it is not as straightforward. Regularizing a model in a conformally invariant way~\cite{Shaposhnikov:2008xi} does not mean that at the quantum level one has conformally invariant theory. The possibility of choosing conformal counter terms and the absence of the trace anomaly should be proven. In the current paper we have presented the proof. From a technical perspective we have shown that conformally invariant counter terms can be chosen allowing to renormalize the theory in the symmetry preserving way. Although we have chosen to consider only scalar fields the results can be generalized to include both spinor and vector fields as well.

One important remark is due here. It is easy to be misleading in saying that the choice of regularization is crucial in understanding the dynamics of the system. It is not so. Choosing one or another regularization might simplify computations. However, the properties of the system should not depend on the regularization chosen. One can not make an arbitrary theory conformally invariant by simply changing the regularization scheme. Of course we do more than that. The system {\it is changed} by adding the dilaton -- additional degree of freedom.

From a physical point of view the proof presents itself a correspondence between low and high energy dynamics of a system with the conformal invariance broken spontaneously. Usually there is no energy scale associated to a conformal field theory. The reason is that usually conformally symmetric vacuum is considered. However, if the spontaneous symmetry breaking occurs certain dimensionful operators acquire non-zero expectation value. This sets the scale in the system. It is natural to expect that at high 
-- compared to the symmetry breaking scale -- energies the system is equivalent to a conformal field theory without any spontaneous symmetry breaking\footnote{It is not necessarily so, see for example~\cite{Dubovsky:2013ira}.}. At the same time the low energy behavior is described by a model with conformal symmetry realized in a non-linear (somewhat hidden) way. The feature of a low energy theory is the presence of the Goldstone mode corresponding to the spontaneous symmetry breaking. Thus, the problem of extending a theory to a conformally invariant one becomes a problem of coupling this theory to the massless dilaton in a consistent way.

We have demonstrated that for theories without Diff anomalies the dilaton stays massless at all orders of perturbation theory. One of the possible extensions of the work is to investigate whether this condition can be relaxed and Diff anomalous theories can be conformal as well.

\section{Acknowledgements}

We are grateful to Riccardo Rattazzi and especially to Mikhail Shaposhnikov for enlightening discussions. The work of A.M. is supported by the Swiss National Science Foundation.

\appendices

\section{Three loop results \label{app_A}}

In this section we present the graphs contributing to the three loop renormalization of the model (\ref{examp_4d}).
The results of the computation are presented in Tables \ref{tab:3L_no_ct} and~\ref{tab:3L_ct}. 

To illustrate how to use the tables we consider as an example one of the contributions to the correlator with 6 $\phi$ legs. Using the Feynman rules and doing the corresponding integrals leads to the following result for the two diagrams
\begin{equation}
\label{eq:ex_table}
\vcenter{\hbox{
\includegraphics[scale=0.8]{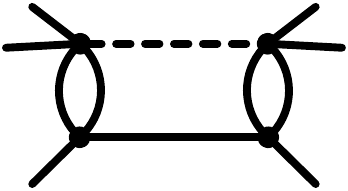} 
}
}
+
\vcenter{\hbox{
\includegraphics[scale=0.8]{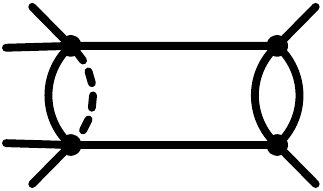} 
}
}
= 
\dfrac{i v ^ {2 \a - 2 }}{
(4\pi)^6 3\eps^3}
(-i\lambda)^2(-i2\lambda \eps)^2
\left\{
\dfrac{
\left[
\dfrac{6!}
{2!^2 2}
\right]
}{4}
+
\dfrac{
\left[
\dfrac{6!}
{2!^2 2}
\right]
}{2}
\right\},
\end{equation}
where solid and dashed lines represent $\phi$ and $\c$ fields respectively. The result can be read off the Table \ref{tab:3L_no_ct}, namely, from the 4th row corresponding to the topology in question. The leading pole structure is presented in the column two. All the combinatorics -- the number of possible contractions and symmetry factors (in curly braces in the equation (\ref{eq:ex_table})) -- and the factors coming from the Feynman rules for the vertices can be found in the column three and four (for correlators with six and eight $\phi$ fields correspondingly).

\begin{table}[H]
\centering
\begin{tabular}
{|c|c|c|c|}
\hline 
Topology
& 
$\eps^{-3}$ part
&
$\phi^6$: $
\dfrac{\lambda^4}{(4\pi)^6} (2\eps)^2
$
&
$\phi^8$:  $
\dfrac{\lambda^4}{(4\pi)^6} (-2\eps)^2
$
\\ \hline
\raisebox{-12pt}{\includegraphics[angle=0]{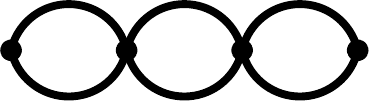}}
&
$
\dfrac{i}{
\eps^3
}
$
&
$
\dfrac{75}{2}
$
&
$
210
$
\\
\hline
\raisebox{-20pt}{\includegraphics[angle=0]{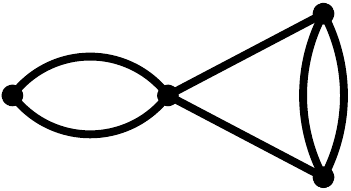}}
&
$\dfrac{i}{
2\eps^3
}$
&
$
\dfrac{195}{2}
$
&
$
175
$
\\
&&&
\\
\hline
%
\raisebox{-20pt}{\includegraphics[angle=0]{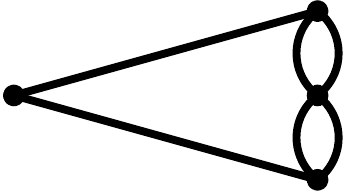}}
&
$
\dfrac{i}{
3\eps^3
}
$
&
$
105
$
&
$
420
$
\\
\hline
%
\raisebox{-20pt}{\includegraphics[angle=0]{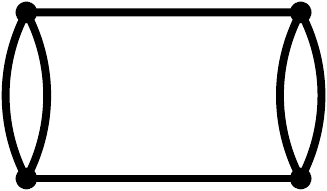}}
&
$
\dfrac{i}{
3\eps^3
}
$
&
$
\dfrac{135}{2}
$
&
$
140
$
\\
\hline
%
%
\raisebox{-20pt}{\includegraphics[angle=0]{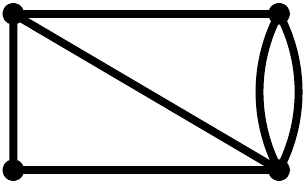}}
&
$
\dfrac{i}{
6\eps^3
}
$
&
$
405
$
&
$
840
$
\\
\hline
%
\raisebox{-20pt}{\includegraphics[angle=0]{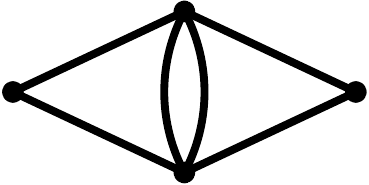}}
&
$
\dfrac{i}{
3\eps^3
}
$
&
$
75
$
&
$
315
$
\\
\hline
%
\raisebox{-12pt}{\includegraphics[angle=0]{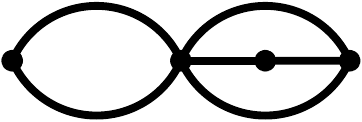}}
&
$
\dfrac{i}{
2\eps^3
}
$
&
$90$
&
$
630
$
\\
\hline
%
\raisebox{-25pt}{\includegraphics[angle=0]{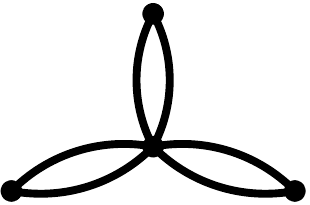}}
&
$
\dfrac{i}{
\eps^3
}
$
&
None
&
$
\dfrac{105}{4}
$
\\
\hline
\end{tabular} 
\caption{\label{tab:3L_no_ct}
Topologies contributing to divergent part of the correlators with six and eight $\phi$ legs with vertices for the tree Lagrangian. }
\end{table}

\begin{table}[H]
\centering
\begin{tabular}{|c|c|c|c|}
\hline
\centering
Topology & $\eps^{-3}$ part &  $\phi^6$: $ \dfrac{\lambda^4}{(4\pi)^6} (2\eps)^2$ & $\phi^8$: $\dfrac{\lambda^4}{(4\pi)^6} (-2\eps)^2 $ 
\\ \hline
\raisebox{-20pt}{\includegraphics[angle=0]{IIL_I.pdf}}
&
$
-\dfrac{3}{2\eps^3
}i
$
&
$
90
$
&
$
315
$
\\
\hline
\raisebox{-20pt}{\includegraphics[angle=0]{IIL_II.pdf}}
&
$
-\dfrac{3}{4\eps^3
}i
$
&
$
225
$
&
$
420
$
\\
\hline
\raisebox{-10pt}{\includegraphics[angle=0]{IL_II_vert.pdf}}
&
$
\dfrac{9}{4\eps^3
}i
$
&
$
30
$
&
$
\dfrac{105}{2}
$
\\
\hline
\end{tabular} 
\caption{\label{tab:3L_ct}
Topologies contributing to divergent part of the correlators with six and eight $\phi$ legs with corresponding counter terms in the vertices.}
\end{table}

%
%
%

\newpage
\bibliographystyle{utphys}
\bibliography{conformal_dilaton}

\end{document}